\documentclass[twocolumn,prb,aps,superscriptaddress,showpacs,amsmath,amssymb,floatfix]{revtex4}

\usepackage{graphicx}
\usepackage{bm}

\begin{document}

\title{Monte Carlo study of thermal fluctuations and Fermi-arc formation in $d$-wave superconductors}

\author{Yong-Wei Zhong}
\affiliation{Department of Physics, Renmin University, Beijing, China}
\affiliation{Tongji Zhejiang College, Jiaxing, Zhejiang, China}

\author{Tao Li}
\affiliation{Department of Physics, Renmin University, Beijing,
China}

\author{Qiang Han}
\affiliation{Department of Physics, Renmin University, Beijing,
China}
\affiliation{Department of Physics and Center of Theoretical
and Computational Physics, The University of Hong Kong, Pokfulam
Road, Hong Kong, China}

\date{\today}

\begin{abstract}
From the perspective of thermal fluctuations, we investigate the pseudogap phenomena in underdoped high-temperature curpate
superconductors. We present a local update Monte Carlo procedure based on the Green's function method to sample the fluctuating
pairing field. The Chebyshev polynomial method is applied to calculate the single-particle spectral function directly and efficiently.
The evolution of Fermi arcs as a function of temperature is studied by examining the spectral function at Fermi energy as well
as the loss of spectral weight. Our results signify the importance of the vortex-like phase fluctuation on the formation of Fermi arcs.
\end{abstract}


\maketitle

\section{Introduction}

The mysterious pseudogap (PG) phase is one of the most fascinating
aspects of the underdoped high-temperature curpate superconductors
(HTCS). By a variety of probes the pseudogap (the suppression of
the low-energy single-particle spectral weight) has been observed to
persist from above the superconducting (SC) critical temperature
$T_{c}$ to $T^*$ in the underdoped regime. The direct evidences of
this spectral gap come from the angle-resolved photoemission
spectroscopy (ARPES) ~\cite{ding, DS, AG}. Ding \textit{et
al.}~\cite{ding} studied the underdoped
Bi$_2$Sr$_2$CaCu$_2$O$_{8+\delta}$ using ARPES and found that a
pseudogap with $d$-wave symmetry begins to open up for $T<T^*$ and
develops smoothly into the $d$-wave SC gap below $T_c$. One peculiar
property of the pseudogap phase revealed by further experimental
investigation is the truncated Fermi surface termed as Fermi
arcs,~\cite{MR} exhibiting distinct difference from the point-like
(four gap nodes) Fermi surface for $T$ well below $T_c$ as expected
for a pure $d$-wave superconductor and the closed Fermi surface for
$T$ above $T^*$.

There are two basic scenarios of the PG phase. The first one
attributes the opening of the pseudogap to the presence of an exotic
order competing with the SC phase, such as the spin \cite{AM, EZ}
and/or charge \cite{CLi} density waves and so on.  The second
scenario associates the PG phase with the phase-incoherent pairing
and therefore the pseudogap is interpreted as a precursor of the SC
order. In this preformed-pair scenario there are two energy scales:
one is the BCS energy gap $\Delta$ which is closely related to the
binding energy of the electron pair, and the other is the
phase-stiffness energy scale $T_{\theta}$ which protects the phase
coherence. For the conventional superconductors $T_{\theta}$ is
larger than $\Delta$ so that the SC state is destroyed by pair
breaking. However for the underdoped HTCS, because of the low
carrier density and the short correlation length, $\Delta$ is larger
than $T_{\theta}$,~\cite{emergy} and therefore the phase coherence
is destroyed while the energy gap survives as temperature increases
across $T_c$. In this context $T_{c}$ is determined by $T_{\theta}$
and the pseudogap is caused by the pair
fluctuations~\cite{emergy,randeria92,randeria98,eckl,mayr,Franz,berg,alvarez,tsvelik,TQ,QTW}.
Franz and Millis~\cite{Franz} showed that random supercurrent
induced by thermal phase fluctuations can cause the shift of
electronic spectral weight in both momentum and energy. Berg and
Altman~\cite{berg} further attributed the emergence of the Fermi arc
to the pile up of the low-energy spectral weight along the
underlying Fermi surface due to the Doppler-shift effect of the
fluctuating supercurrent. This picture of phase fluctuations is
concise and instructive, yet the analytical results relied on the
semiclassical approximation \cite{Franz} where only far-field effect
of the vortex-type excitations is considered, which might be
uncontrolled as argued in Ref.[\onlinecite{tsvelik}]. Furthermore,
the probability distribution of the fluctuating supercurrent was
assumed phenomenologically to be Gaussian type. Recently we
\cite{TQ,QTW} attempted to go beyond the semiclassical approximation
by employing a 2D XY model to simulate the vortex-type phase
fluctuations and numerically taking both the Doppler effect of the
whirling supercurrent and the scattering effect of vortices as
topological singularities into full consideration. However, the XY
model is still a phenomenological description of the phase
fluctuations, which includes a temperature-independent
phase-stiffness constant $J$.

In this work,
we start from a 2D attractive Hubbard model with only nearest-neighbor interactions to
investigate the pseudogap phase and the evolution of Fermi arcs in $d$-wave superconductors.
The path-integral formalism is employed where pairing fluctuations are inherently embedded.
A local-update Monte Carlo scheme on the basis
of the Green's function method is presented to speed up the random walk
in the classical configuration space of pairing field. Superfluid
density is calculated as the signature of the SC phase transition
and compared with the phase correlation function~\cite{mayr}. The
single-electron spectral function is calculated using Chebyshev
polynomial method~\cite{weibe,erez,TQ,QTW}. The temperature dependence of Fermi-arc length is
discussed.

The paper is organized as follows: In Section \ref{model}
we describe the basic path-integral formalism to treat the 2D
extended Hubbard model. The local-update algorithm based on Green's
function theory and the Chebyshev expansion approach are presented.
In Section \ref{results}, we calculate the temperature dependencies
of various quantities relevant to the phase fluctuations and
pseudogap phase. The conclusion is given in Section
\ref{conclusion}.

\section{The model and formalism}
\label{model}

The BCS Hamiltonian $ \hat{H}_\text{BCS} $ we adopt is given by:
\begin{eqnarray}
\hat{H}_\text{BCS} & = & -t\sum_{i,\delta,\sigma}c_{i\sigma}^\dagger c_{i+\delta\sigma}-t^\prime\sum_{i,\delta^\prime,\sigma}c_{i\sigma}^\dagger c_{i+\delta^\prime\sigma} \nonumber \\
          & & -V\sum_{i,\delta}c_{i\uparrow}^\dagger c_{i\uparrow}c_{i+\delta\downarrow}^\dagger c_{i+\delta\downarrow}-\mu\sum_{i,\sigma}n_{i,\sigma}.
\end{eqnarray}
Here $\sigma$ denotes spin. $i$ is the index of site of the two-dimensional $L\times L$ square lattice.
$i+\delta$ and $i+\delta^\prime$ denote the nearest-neighboring (NN) and next-NN sites of $i$, respectively.
$t$ and $t^\prime$ are the NN and next-NN hopping integrals, respectively. $\mu$ is the chemical potential.
The attractive interaction $V>0$ between electrons on the NN sites favors unconventional superconducting phases.
To investigate the effect of superconducting fluctuations, the quantum partition function is expressed in the
path integral formalism ~\cite{field,phasenature}:
\begin{eqnarray}
Z & = & \int D \{ \varphi_{i\sigma}(\tau),\bar{\varphi}_{i\sigma}(\tau) \} \exp(-S)
\end{eqnarray}
where the action $S$ is expressed as
\begin{widetext}
\begin{equation}
S(\varphi,\bar{\varphi})  = \int_{0}^{\beta}d\tau \left[\sum_{i\sigma}\bar{\varphi}_{i\sigma}(\tau)(\partial_{\tau}-\mu)\varphi_{i\sigma}(\tau)
-\sum_{i,j,\sigma} t_{ij} \bar{\varphi}_{i\sigma}(\tau) \varphi_{j\sigma}(\tau)
-V\sum_{i,\delta}\bar{\varphi}_{i\uparrow}(\tau) \bar{\varphi}_{i+\delta\downarrow}(\tau) \varphi_{i+\delta\downarrow}(\tau) \varphi_{i\uparrow}(\tau) \right],
\end{equation}
\end{widetext}
where $\varphi_{i\sigma}$ and $\bar{\varphi}_{i\sigma}$ denote Grassmann fields and $\beta=1/k_{B}T$.
We then decouple the quartic term in the action
by introducing an auxiliary Hubbard-Stratonovich field $\Delta_{i,i+\delta}(\tau)$ in the Cooper channel.
For a square lattice with $N$ sites and  periodic boundary condition, there are totally $2N(N=L^2)$
independent $\Delta_{i,i+\delta}(\tau)$'s. Hereafter we use $\Delta$ to denote the
set $\{\Delta_{i,i+\delta}(\tau)\}$. The partition function now becomes
\begin{equation}
Z = \int D\Delta D\bar{\Delta} e^{-\beta\Omega(\Delta,\bar{\Delta})}, \label{partfunc}
\end{equation}
where
\begin{equation}
\int D\Delta D\bar{\Delta} \equiv \int \prod_{i=1}^N \prod_{\delta=\hat{x},\hat{y}} d\Delta_{i,i+\delta}(\tau) d\bar{\Delta}_{i,i+\delta}(\tau).
\end{equation}
In Eq.~(\ref{partfunc}), the grand potential is expressed as
\begin{equation}
\Omega(\Delta,\bar{\Delta})= \Omega_f(\Delta,\bar{\Delta}) + V^{-1} \sum_{i,\delta}|\Delta_{i,i+\delta}(\tau)|^2,
\end{equation}
where $\Omega_f$ denotes the fermionic thermodynamic potential
\begin{equation}
\Omega_f(\Delta,\bar{\Delta}) = -\beta^{-1} \ln \text{Tr} e^{-\beta \hat{H}_\text{BdG}(\Delta)}.
 \end{equation}
Here the BdG Hamiltonian is written by
\begin{eqnarray}
\hat{H}_\text{BdG}(\Delta) &=& \Psi^\dagger \tilde{H}_\text{BdG}(\Delta) \Psi \label{bdgmtx} \\
&=&
\sum_{i,j=1}^N
( c_{i\uparrow}^\dagger, c_{i\downarrow} )
\left(
\begin{array}{cc}
-t_{i,j} & \Delta_{i,j}\\
\Delta_{i,j}^* & t_{i,j}
\end{array}
\right)
\left(
\begin{array}{c}
c_{j\uparrow} \\
c_{j\downarrow}^\dagger
\end{array}
\right)
\nonumber
\end{eqnarray}
where $\Psi^\dagger$($\Psi$) denotes the Nambu creation (annihilation) operator defined as
$\Psi^\dagger=(c_{1\uparrow}^\dagger,c_{1\downarrow}, c_{2\uparrow}^\dagger,c_{2\downarrow},
\cdots, c_{N\uparrow}^\dagger,c_{N\downarrow})$. $\tilde{H}_\text{BdG}$ is a $2N\times 2N$
Hermitian matrix which will be called BdG matrix. Hereafter we use capital letters with
a tilde ( $\tilde{}$ ) to denote $2N\times 2N$ dimensional matrices (e.g. the BdG matrix $\tilde{H}_\text{BdG}$) while a hat
( $\hat{}$ ) to denote operators in second quantization (e.g. the BdG Hamiltonian $\hat{H}_\text{BdG}$).
For the sake of convenience, we will omit the argument $\Delta$ and simply use $\hat{H}_\text{BdG}$ and $\tilde{H}_\text{BdG}$
to denote the BdG Hamiltonian and matrix for a certain pairing field $\Delta$.

In the following, we will ignore the $\tau$-dependence of $\Delta_{i,i+\delta}(\tau)$, i.e.
the quantum fluctuation, and concentrate on its thermal fluctuations expected to be dominant near
$T_c$ especially in the high temperature pseudogap region.  With this approximation, $\hat{H}_\text{BdG}$
actually describes the electrons moving in a static but spatially fluctuating pairing field.  Moreover
Eq.~(\ref{partfunc}) becomes a classical partition function expressed as an integration over the classical
phase space formed by $\{\Delta_{i,i+\delta}=|\Delta_{i,i+\delta}|e^{i\phi_{i}^{\delta}}\}$,
whose dimension is $4N$ for a $N$-site square lattice. Such multidimensional integration can be performed
by the standard Monte Carlo method. To achieve this goal, one need to obtain the probability distribution
$P(\Delta)\propto e^{-\beta\Omega(\Delta,\bar{\Delta})}$ for a configuration $\Delta$, or its change characterized by the ratio
\begin{equation}
\frac{P(\Delta^\prime)}{P(\Delta)}=e^{-\beta [\Omega(\Delta^\prime,\bar{\Delta}^\prime)-\Omega(\Delta,\bar{\Delta})]},
\end{equation}
for a possible change of configuration $\Delta\rightarrow \Delta^\prime$. The acceptance probability for such a change is given by
\begin{equation}
P_\text{A}(\Delta^\prime\leftarrow \Delta)=\min\left[ 1, \frac{P(\Delta^\prime)}{P(\Delta)} \right]
\end{equation}
according to the Metropolis algorithm. To obtain $\Omega$ especially the nontrivial $\Omega_f$,
previous numerical work~\cite{mayr} related it to the eigen-spectrum of the BdG matrix through
the relation $\Omega_f(\Delta,\bar{\Delta})=-\beta^{-1}\sum_n \ln(1+e^{-\beta\epsilon_n})$, where
$\epsilon_n$ is the eigenvalue of the BdG equations,
\begin{equation}
\sum_{j}
\left(
\begin{array}{cc}
-t_{ij} & \Delta_{i,j} \\
\Delta_{i,j}^* & t_{i,j}^*
\end{array}
\right)
\left(
\begin{array}{c}
u_n^j \\
v_n^j
\end{array}
\right)
=
\epsilon_n
\left(
\begin{array}{c}
u_n^i \\
v_n^i
\end{array}
\right)
\label{bdg}
\end{equation}
To solve this eigenvalue problem, one needs to diagonalize a $2N\times 2N$ BdG matrix and the workload is $O(N^3)$
which is quite time-consuming for large lattice.

In this paper, we will propose an alternative local-update scheme based on the Green's function method.
The Gor'kov Green's function is employed,
which is defined as:
\begin{equation}
G(i\tau,j\tau^\prime)  = -\text{T}
\left\langle
          \begin{array}{cc}
                    c_{i\uparrow}(\tau)c_{j\uparrow}^\dagger(\tau^\prime) & c_{i\uparrow}(\tau)c_{j\downarrow}(\tau^\prime)\\
                    c_{i\downarrow}^\dagger(\tau) c_{j\uparrow}^\dagger(\tau^\prime) & c_{i\downarrow}^\dagger(\tau) c_{j\downarrow}(\tau^\prime)
          \end{array}
      \right\rangle,
\end{equation}
in terms of $2\times 2$ Nambu matrix notation. Here $\langle\cdots\rangle=\text{Tr}[\cdots e^{-\beta \hat{H}_\text{BdG}}]
/\text{Tr}[e^{-\beta \hat{H}_\text{BdG}}]$,
and its Fourier transform with respect to the imaginary time is
\begin{equation}
G(i,j; i\omega_k) = \int_0^\beta d\tau G(i\tau,j0) e^{i\omega_k \tau}
\end{equation}
where $\omega_k=2\pi T(k+1)$ the Matsubara frequencies. One can easily find that the matrix form of the Gor'kov Green's
function is actually the resolvent of the BdG matrix $\tilde{H}_\text{BdG}$, i.e.
\begin{equation}
\tilde{G}(i\omega_k) = (i\omega_k \tilde{I} - \tilde{H}_\text{BdG})^{-1}, \label{res}
\end{equation}
where $\tilde{I}$ the unity matrix. Combining this relation with Eq.~(\ref{bdg}), we obtain the spectral
representation of $\tilde{G}(i\omega_k)$,
\begin{equation}
G(i,j;i\omega_k) = \sum_{\epsilon_n} \frac
{
\left(
\begin{array}{c}
u_n^i  \\
v_n^i
\end{array}
\right)
(u_n^j,v_n^j)^*
}
{i\omega_k - \epsilon_n}
\label{specrep}
\end{equation}
which is a $2\times 2$ matrix.
Before starting our simulation, we only need to diagonalize the BdG matrix with certain initial (often random)
configuration once for all. Then the eigen-energies $\epsilon_n$ and eigen-functions $u_n$ and $v_n$  are used to
calculate the Green's function according to Eq.~(\ref{specrep}). As we will show in Section \ref{algorithm},
we can always update the Gor'kov Green's function without having to diagonalize the BdG matrix any more as long
as the change of configuration is proposed locally.

\subsection{Update of the Green's function and calculation of the acceptance probability}
\label{algorithm}

We assume a local change of the configuration, located at the 1 and 2 sites without loss of generality, from
$\Delta_{1,2}$ to $\Delta_{1,2}^\prime = \Delta_{1,2} + \chi_{1,2}$. Then the BdG Hamiltonian becomes
\begin{eqnarray}
&& \hat{H}_\text{BdG}^\prime = \hat{H}_\text{BdG} + \hat{H}_1 \\
&& \hat{H}_1 = \chi_{1,2} (c_{1\uparrow}^\dagger c_{2\downarrow}^\dagger + c_{2\uparrow}^\dagger c_{1\downarrow}^\dagger) + h.c.
\end{eqnarray}
where $\hat{H}_1$ denotes the corresponding change of the BdG Hamiltonian.  According to Eq.~(\ref{res}), we have
\begin{eqnarray}
\tilde{G}^\prime(i\omega_n) &=&  (i\omega_k \tilde{I} - \tilde{H}_\text{BdG} - \tilde{H}_1 )^{-1} \nonumber \\
&=& \tilde{G}(i\omega_n) [1- \tilde{H}_1 \tilde{G}(i\omega_n)]^{-1} ,
\label{update}
\end{eqnarray}
where $\tilde{H}_1$ denotes the matrix form of $\hat{H}_1$ in the Nambu representation as Eq.(\ref{bdgmtx}),
\begin{equation}
\tilde{H}_1 = \left(
\begin{array}{ccc}
X_{4\times 4} & \vline & 0 \\
\hline
0 & \vline & 0
\end{array}
\right)_{2N\times 2N}
\end{equation}
where only its upper-left-corner $4\times 4$ block is has non-zero elements, and $X_{4\times4}$ is
\begin{equation}
X_{4\times4} = \left(
\begin{array}{cccc}
0               &   0           &   0               &   \chi_{1,2} \\
0               &   0           &   \chi_{1,2}^*    &   0 \\
0               &   \chi_{1,2}  &   0               &   0 \\
\chi_{1,2}^*    &   0           &   0               &   0
\end{array}
\right).
\end{equation}
The inverse operation on the right hand of Eq.~(\ref{update}) can be performed as follows,
\begin{eqnarray}
&& (I-\tilde{H}_1 \tilde{G})^{-1} = \left[ I-
\left(
\begin{array}{cc}
X & 0 \\
0 & 0
\end{array}
\right)
\left(
\begin{array}{cc}
A & B \\
C & D
\end{array}
\right) \right]^{-1} \\
&=&
\left(
\begin{array}{cc}
I-XA & -XB \\
0 & I
\end{array}
\right)^{-1} \\
&=&
\left[
\begin{array}{cc}
(I-XA)^{-1} & (I-XA)^{-1} XB \\
0 & I
\end{array}
\right]. \label{detail}
\end{eqnarray}
In the above derivation, we use block matrices $A, B, C, D$ to denote $\tilde{G}$, whose dimension is
$4\times 4$, $4\times (2N-4)$, $(2N-4)\times 4$ and $(2N-4)\times (2N-4)$, respectively.
As most non-zero elements of the above matrix is concentrated on the first four rows, one can update
$\tilde{G}^\prime$ according to Eq.~(\ref{update}) with $O(N^2)$ computing operations.

Next, we will shown how to obtain the change of the thermodynamic potential which determines the acceptance
of the proposed local update. According to textbook~\cite{no}, one has
\begin{equation}
\Omega_f^{\prime}-\Omega_f=\int_{0}^{1}d\lambda \langle \hat{H}_1 \rangle_{\lambda}
\label{gp}
\end{equation}
where $\langle\cdots\rangle_{\lambda}=\text{Tr}[\cdots e^{-\beta \hat{H}(\lambda)}]/\text{Tr}[e^{-\beta \hat{H}(\lambda)}]$
with $\hat{H}(\lambda)=\hat{H}_\text{BdG}+\lambda \hat{H}_1$. The integrated function $\langle \hat{H}_1 \rangle_{\lambda}$
are also related with the Gor'kov Green's function \cite{no}:
\begin{eqnarray}
&&\langle \hat{H}_1 \rangle_{\lambda} = 2\text{Re}\{ \chi_{12} [ G_{\lambda}(1\tau,2\tau^{+})_{21} + G_{\lambda}(2\tau,1\tau^{+})_{21} ] \} \nonumber \\
&&= 2\text{Re} \{ \chi_{12} T \sum_{i\omega_k}[ G_{\lambda}(1,2;i\omega_k)_{21} + G_{\lambda}(2,1;i\omega_k)_{21} ] \}  \nonumber \\
 \label{h1a}
\end{eqnarray}
Using Eq.~(\ref{update}) and (\ref{detail}), we have
\begin{eqnarray}
G_{\lambda}(1,2;i\omega_k)_{21} &=& [G(i\omega_k)_{4 \times 4}(I-\lambda XA)^{-1}]_{2,3} \label{g12}\\
G_{\lambda}(2,1;i\omega_k)_{21} &=& [G(i\omega_k)_{4 \times 4}(I-\lambda XA)^{-1}]_{4,1}. \label{g21}
\end{eqnarray}
Therefore, from Eqs.~(\ref{h1a}), (\ref{g12}) and (\ref{g21}), the integration over $\lambda$ in Eq.~(\ref{gp}) is readily performed,
\begin{eqnarray}
&& \int_0^1 (I-\lambda XA)^{-1} d\lambda  = -(XA)^{-1} \ln ( I - XA ) \label{ln} \\
&=& -O
\left(
\begin{array}{cccc}
\frac{\ln (1-d_1)}{d1} & 0 & 0 &  0 \\
0 & \frac{\ln (1-d_2)}{d_2} & 0 & 0 \\
0 & 0 & \frac{\ln (1-d_3)}{d_3} & 0 \\
0 & 0 & 0 & \frac{\ln (1-d_4)}{d_4}
\end{array}
\right)
O^{-1}, \nonumber
\end{eqnarray}
where to treat the $4\times 4$ matrix as an argument of logarithm function in Eq.~(\ref{ln}), we diagonalize $XA=ODO^{-1}$ with $O$ the
transformation matrix and $D$ the diagonal matrix with eigenvalues $d_{1,2,3,4}$.

\subsection{Spectral function and Chebyshev polynomial approach}

With the help of the local-update scheme described above, the classical phase space of $\Delta$ are sampled and thermodynamic
averages of physical
quantities can be obtained. As an example, we give the definition of the single-electron spectral function,
\begin{equation}
A(\mathbf{k},\omega) = \frac{ \int D\Delta D\bar{\Delta} A_\Delta(\mathbf{k},\omega) e^{-\beta\Omega(\Delta,\bar{\Delta})} }
{\int D\Delta D\bar{\Delta} e^{-\beta\Omega(\Delta,\bar{\Delta})}}
\end{equation}
where $A_\Delta(\mathbf{k},\omega)$ denotes the single-electron spectral function for a certain configuration of
$\Delta$. $A_\Delta(\mathbf{k},\omega)$ can be derived according to
\begin{eqnarray}
A_\Delta(\mathbf{k},\omega) &=& -\frac{1}{\pi}\text{Im}\sum_{i,j} G^\text{R}(i,j,\omega)_{11} e^{i\mathbf{k}\cdot(\mathbf{i}-\mathbf{j})},
\end{eqnarray}
where the retarded Green's function is the real-space representation of the resolvent,
\begin{equation}
G^\text{R}(i,j,\omega)_{11} = \langle i\uparrow| (\omega+i0^+ - \hat{H}_\text{BdG})^{-1} |j\uparrow \rangle.
\end{equation}
Combining the above two equations, we have
\begin{equation}
A_\Delta(\mathbf{k},\omega) =  \langle \mathbf{k} \uparrow| \delta(\omega-\hat{H}_\text{BdG}) |\mathbf{k}\uparrow \rangle,
\label{spec}
\end{equation}
where $|\mathbf{k}\uparrow\rangle = N^{-1/2} \sum_i e^{i\mathbf{k}\cdot\mathbf{i}} c_{i\uparrow}^\dagger |0\rangle$.
Generally, one can first solve
the BdG equation (\ref{bdg}), then employ the following equation,
\begin{equation}
A_\Delta(\mathbf{k},\omega) = \sum_{\epsilon_n,i,j} \delta(\omega - \epsilon_n) u_n^i u_n^{j*} e^{i\mathbf{k}\cdot (\mathbf{i-j})}
\end{equation}
to calculate the spectral function for one configuration. However, since the computational effort of full diagonalization of
the BdG matrix is $O(N^3)$, we will apply the Chebyshev polynomial approach \cite{weibe,QTW}, which is $O(MN)$ with $M\ll N^2$,
to cut the computational cost.

We perform a Chebyshev polynomial expansion
\begin{equation}
\delta(x-y) = \frac{1}{\pi\sqrt{1-x^2}} \left[\mu_0 + 2\sum_{m=1}^\infty T_m(x) T_m(y) \right], \label{deltafunc}
\end{equation}
to handle the Dirac $\delta$ function. After substituting it into Eq~(\ref{spec}). we have
\begin{equation}
A_\Delta(\mathbf{k},\omega) = \frac{\mu_0 + 2\sum_{m=1}^M \mu_m g_m T_m(\omega/s)}{\pi\sqrt{1-(\omega/s)^2}}, \label{cheby}
\end{equation}
where
\begin{equation}
\mu_m=\langle \mathbf{k} \uparrow| T_m(\hat{H}_\text{BdG}/s) |\mathbf{k}\uparrow \rangle, \label{moments}
\end{equation}
are Chebyshev moments. Here for numerical calculation, the infinite series in Eq.~(\ref{deltafunc}) has to be truncated by
$M$ as shown in Eq.~(\ref{cheby}) and to damp the consequential Gibbs oscillations the Lorentz kernel $g_m$ is used in Eq.~(\ref{cheby})
with
$g_m = \sinh[\lambda(1-m/M)]/\sinh(\lambda)$ where $\lambda$ is a free parameter of the kernel and we choose
$\lambda=4$ throughout our calculation as a compromise between good resolution and sufficient damping of the Gibbs
oscillations as suggested in Ref.~[\onlinecite{weibe}]. $s$ denotes the scaling factor ensuring the spectrum of
$\hat{H}_\text{BdG}/s$ falling into the interval $[-1,1]$, i.e. the domain of the Chebyshev polynomials.

Most computational effort is spent in the calculation of the Chebyshev moments $\mu_m$ according to Eq.~(\ref{moments}),
which reduces to sparse matrix-vector multiplications after taking advantage of the recursion relation $T_m(x)=2xT_{m-1}(x)-T_{m-2}(x)$.
Considering that the BdG Hamiltonian is sparse, the cost of matrix-vector multiplication is an $O(N)$ process and the calculation of $M$
moments requires only $O(MN)$ computational operations. Further relations of the Chebyshev polynomials
$T_{2m}=2T_m^2-1$ and $T_{2m+1}=2T_mT_{m+1}-T_1$ enable us to obtain two moments per matrix-vector multiplication.
Therefore, calculation of the single-particle spectral function using the Chebyshev polynomial method is fast, efficient
and direct with less memory consuming, superior to direct diagonalization [generally $O(N^3)$] of the BdG matrix.

\begin{figure*}[ht]
\begin{tabular}{ll}
\includegraphics[width=6.5cm]{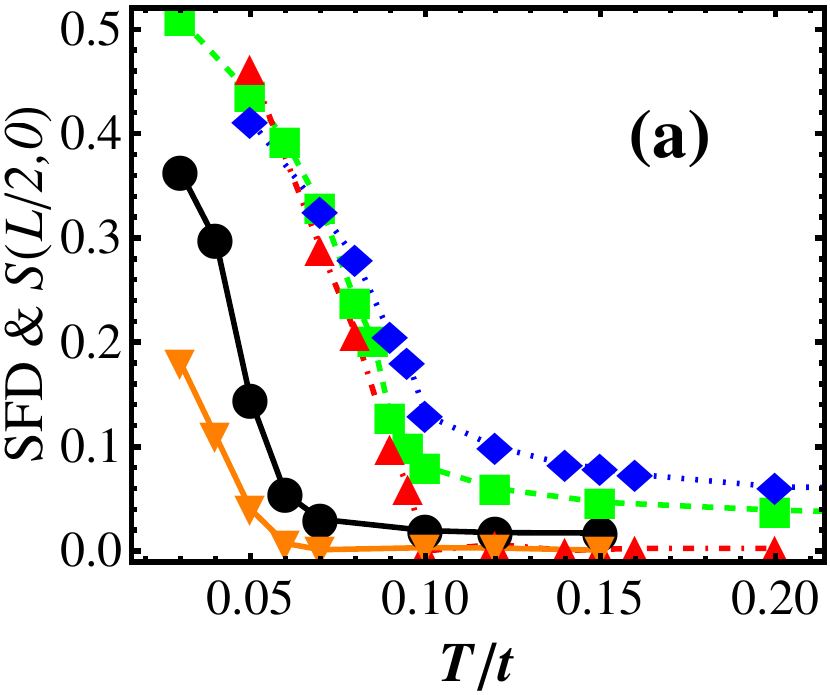} & \includegraphics[width=6.5cm]{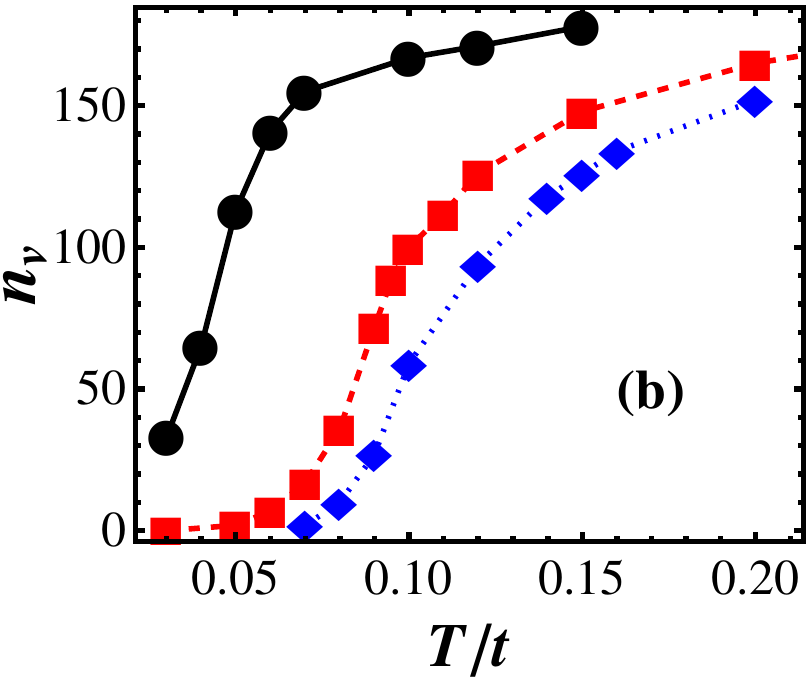} \\
\includegraphics[width=8.5cm]{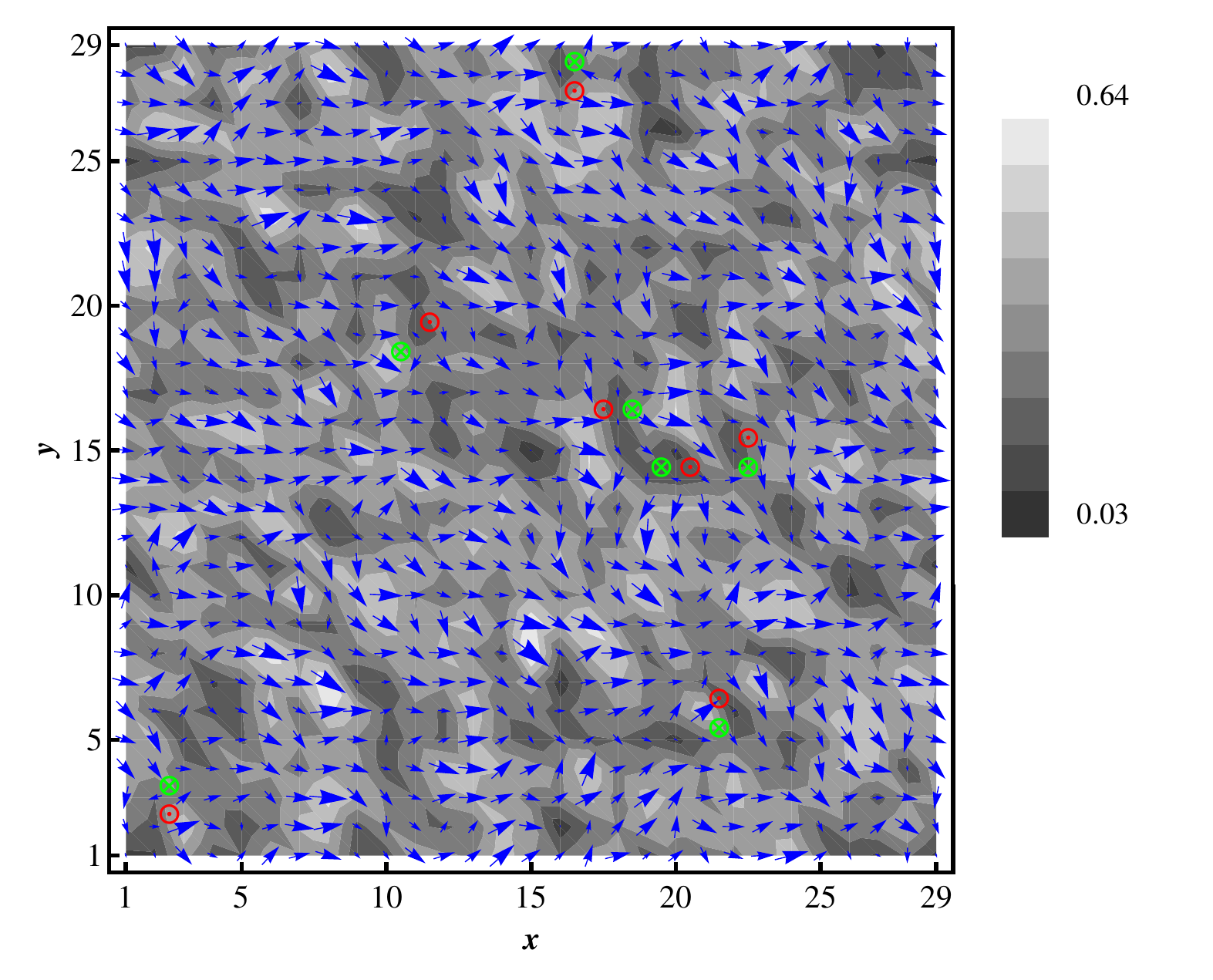} & \includegraphics[width=8.5cm]{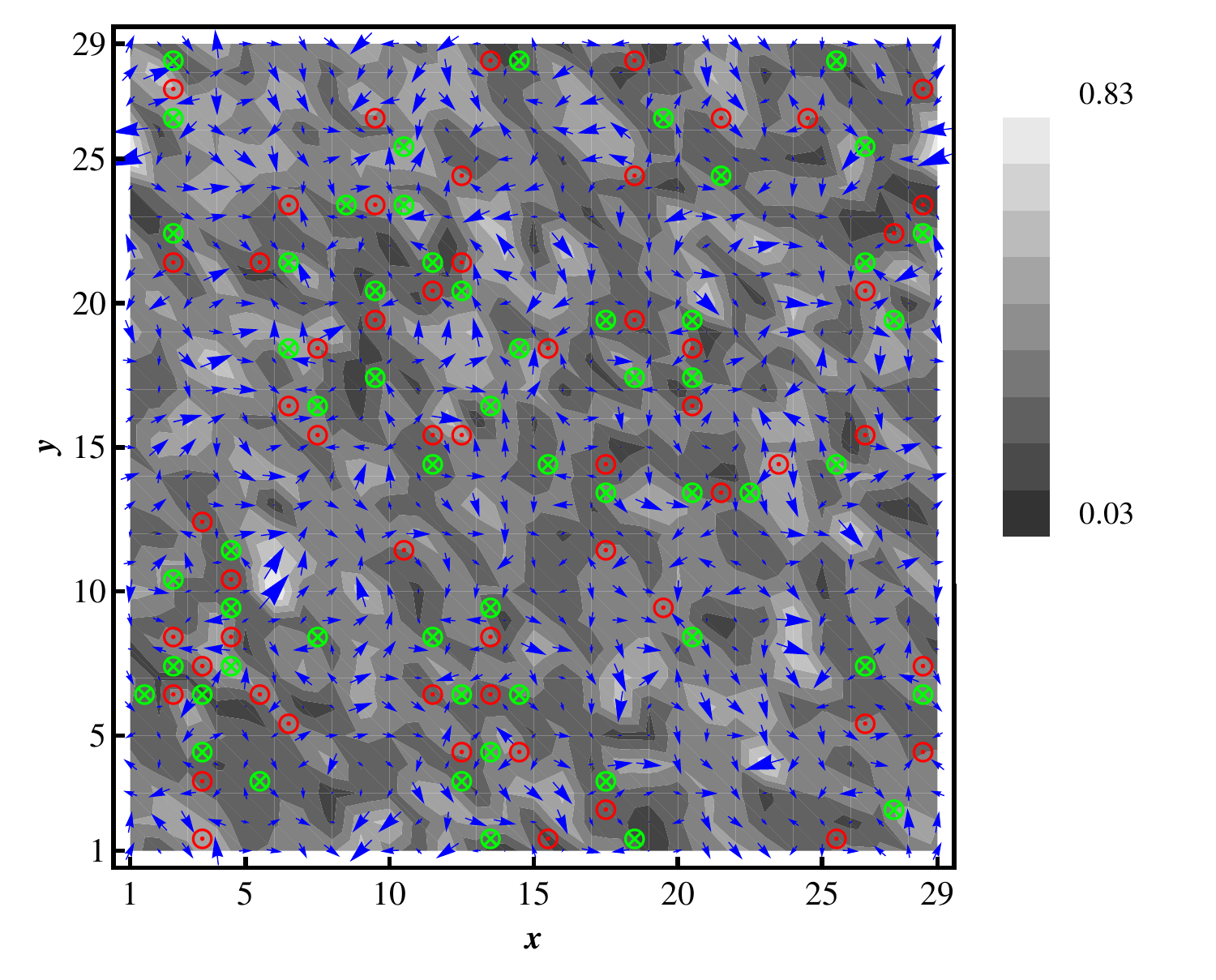}
\end{tabular}
\caption{
Upper-left panel: Superfluid density(SFD) and the long-range phase
correlation function $s(14,0)$ vs $T$ for three different
interactions. Black disks (SFD, $V=1.2$), green
square (SFD, $V=2.4$), blue
diamond (SFD, $V=4.0$), orange down-triangle (S(14,0), $V=1.2$), and red triangle (S(14,0), $V=4.0$). $T_{c}$
is around $0.05$, $0.09$, $0.105$ respectively. Upper-right panel: The $d$-wave vortex number $n_v$ vs $T$ for three
different interactions. For each interaction, there is a region that
vortices gush abruptly. Black disks($V=1.2$), red square($V=2.4$), and
blue diamond($V=4.0$).
Panels of the Lower row: Snap shots of the spatial distribution of the magnitude and phase of the $d$-wave order parameter
$\Delta_d(i)$ on a $29\times29$ lattice (note the periodic boundary condition) at $T=0.06$ (lower left) and $T=0.12$ (lower right).
The phase on each lattice site is represented by a blue arrow, while the magnitude is represented by the size of
the arrow as well as the gray scale density. Also shown are the topological excitations with the red $\bigodot$
denoting the vortex with winding number 1 and green $\bigotimes$ denoting the antivortex with winding number $-1$.
}
\label{vortices}
\end{figure*}

\section{Numerical results}
\label{results}

In our work, the parameters are chosen as below: $t=1$ (as the unit of energy), $t^\prime=-0.3$, $\mu=-0.83$, and $N=L\times L=28\times 28$. According to the above parameters the average electron number is approximately $0.9$ and accordingly the hole doping is $0.1$. For each temperature,
the first $10^3$ MC sweeps are dropped to equilibrate the system. $10^3$ configurations are used as samples to get statistical average. Each MC sweep includes $2N^2$ local updates to reduce the configuration correlation. By these arguments, the statistical error reduces to an
acceptable level. Then, we calculate the spectral function directly with the help of the Chebyshev polynomials~\cite{weibe} with the
truncation $M=2048$.

In Fig.~\ref{vortices}(a), we show the superfluid density (SFD) $D_{s}/\pi e^2$ (see Appendix) as a function of temperature for different pairing interactions. The SFD decreases as the temperature increases and displays an apparent drop indicating a phase transition for the three interaction strengths, although SFD has a long tail above the transition temperature due to the finite size effect. By taking the leading-edge midpoint as $T_{c}$, we have $T_{c}\approx 0.05, 0.09, 0.105$ for $V=1.2, 2.4, 4$, respectively . Compared with the previous work \cite{mayr},  we find that the transition temperature
increases with the interaction strength at least for $V\leq 4$. This increase of $T_c$ with $V$ for small to intermediate value of $V$ is further supported if we further examine the long-range phase correlation\cite{mayr} $S(L/2,0)=\frac{1}{N}
\sum_{i=1}^N<e^{i\phi_{i}^{\widehat{x}}}e^{-i\phi_{i+(L/2,0)}^{\widehat{x}}}>$, where $\phi_{i}^{\widehat{x}}$ denotes the
phase of the pairing field $\Delta(i,i+\hat{x})$.
The results are also shown in Fig.~\ref{vortices}(a), indicating that both the SFD and $S(L/2,0)$ are measures of the phase stiffness of the condensate.

Conventionally, the phase fluctuation scenario relates the normal to SC phase
transition in underdoped HTCS to the KT-type phase transition. Although
people have made lots of efforts, it is still unclear how this
occurs. Generally, fluctuations of the phase degrees of freedom of the
superconducting order parameter can be described by the 2D XY model, the studies of which have revealed that
the proliferation and unbinding of the vortex-antivortex pairs above $T_\text{KT}$ destroys the
quasi-long-range phase coherence. Here we explore this aspect by observing
the vortex-type excitations in the phase field $\varphi_d(i)$ of the $d$-wave order parameter $\Delta_d(i)=|\Delta_d(i)|e^{i\varphi_d(i)}$.
The definition of $\Delta_d(i)$ is $\Delta_d(i)\equiv[\Delta(i,i+\hat{x})+\Delta(i,i-\hat{x})-\Delta(i,i+\hat{y})-\Delta(i,i-\hat{y})]/4$.
The (anti)vortices are plaquette-centered topological defects of the phase field.
The winding number or vorticity of the (anti)vortex is defined as the anticlockwise sum of the phase
difference around each plaquette of the square lattice (divided by $2\pi$).
For each plaquette labeled by its lower-left corner $i$, its four vertices anticlockwisely
are $i_1=i,i_2=i+\hat{x},i_3=i+\hat{x}+\hat{y},i_4=i+\hat{y}$. The
phase difference between two NN sites for instance $i_2$ and $i_1$
is $\theta_{2,1}(i)\equiv\varphi_d(i_2)-\varphi_d(i_1)=\text{Im}\log(\Delta_d(i_2)\Delta_d^*(i_1))$.
Therefore the winding number around the plaquette $i$ is
$w(i)=[\theta_{2,1}(i)+\theta_{3,2}(i)+\theta_{4,3}(i)+\theta_{1,4}(i)]/2\pi$.
$w(i)=1$ or $-1$ represents a vortex- or antivortex-type topological defect around the plaquette $i$.
The total number of (anti)vortices, i.e.~$n_v$, which quantifies
the phase fluctuation relevant to the topological excitations,
is calculated according to $n_v=\sum_{i} \delta_{w(i),1}$, i.e.~the total number of plaquettes with $w(i)=1$ (note that without external magnetic field the antivortex number always equals to the vortex number).
The temperature dependence of $n_v$ is shown in Figs.~\ref{vortices}(b). One can observe the abrupt jump of $n_v$
at approximately the same temperature obtained from Fig.~\ref{vortices}(a), giving further indication of the KT-type phase
transition.
For illustration, Figs.~\ref{vortices}(c) and (d) show snapshots of the pairing fields recorded at $T=0.06$ and $T=0.12$ for $V=2.4$.
For these two temperatures the coherence length of the $d$-wave order parameter is small and the vortices can be clearly identified
as shown in the figures.
At the temperature well below $T_\text{KT}$, the
vortex density is dilute and all vortices are bound together as vortex-antivortex pairs as shown in Fig.~\ref{vortices}(c).
Above $T_\text{KT}$, vortices gush and the unbinding of vortex-antivortex pair is clearly illustrated as shown in Fig.~\ref{vortices}(d). This behavior conforms to the characteristics of KT phase transition.  Together with the observation that the crossover regions of the SFD and phase correlation reduplicate that of the vortices, we can argue that the SC phase transition is of the KT type.

\begin{figure}[ht]
\begin{tabular}{cc}
\includegraphics[width=4.25cm]{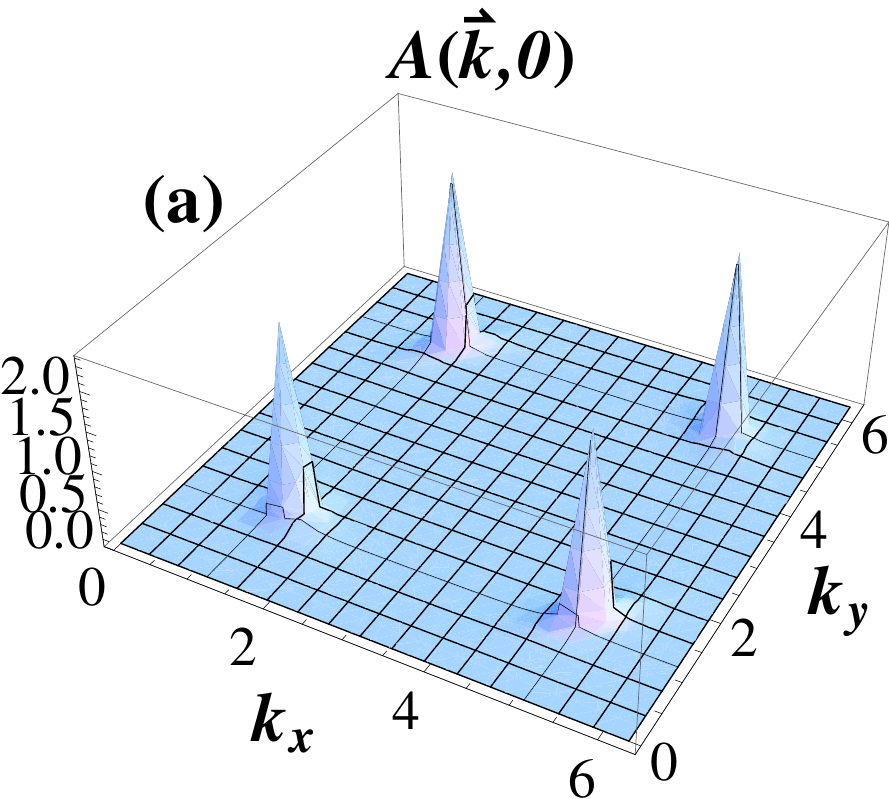} & \includegraphics[width=4.25cm]{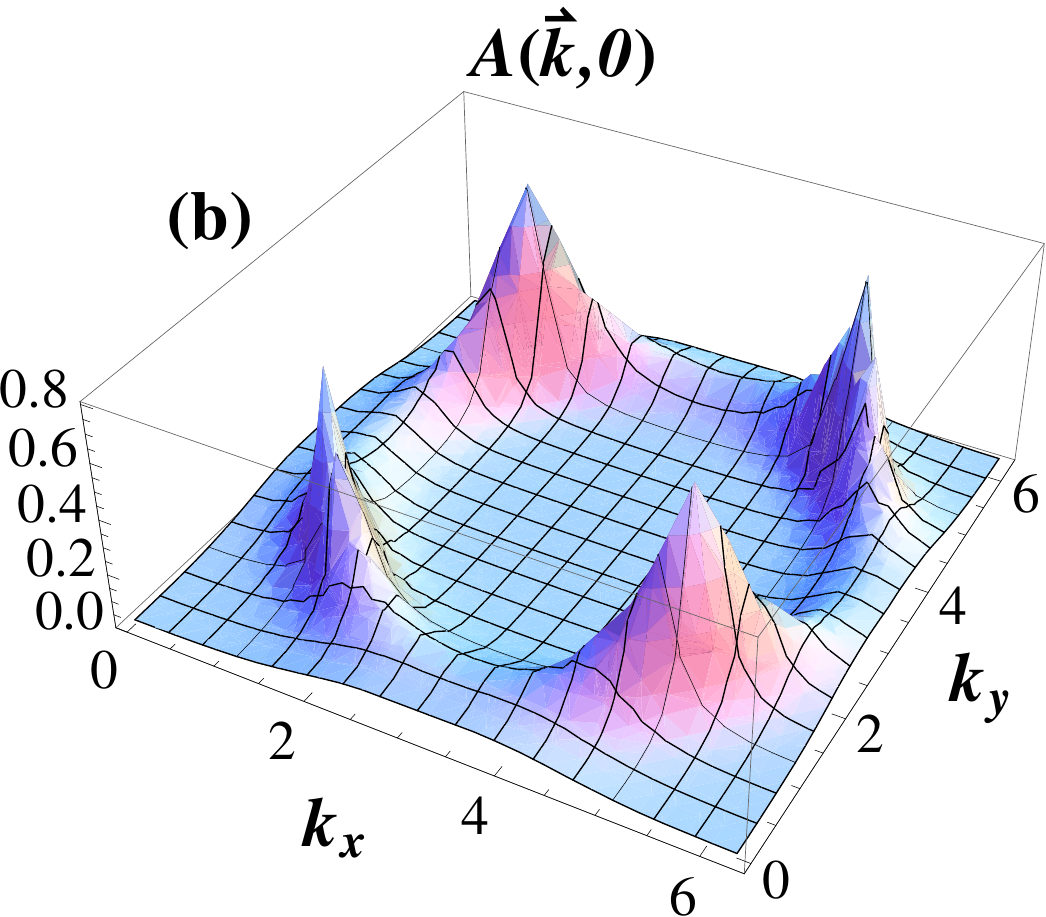} \\
\includegraphics[width=4.25cm]{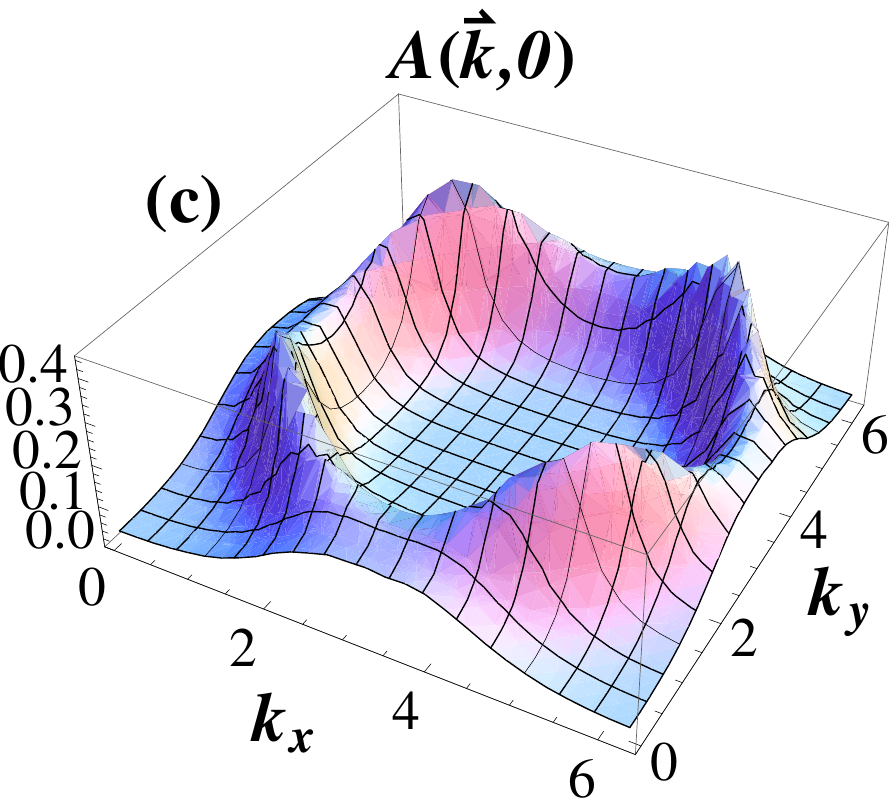} & \includegraphics[width=4.25cm]{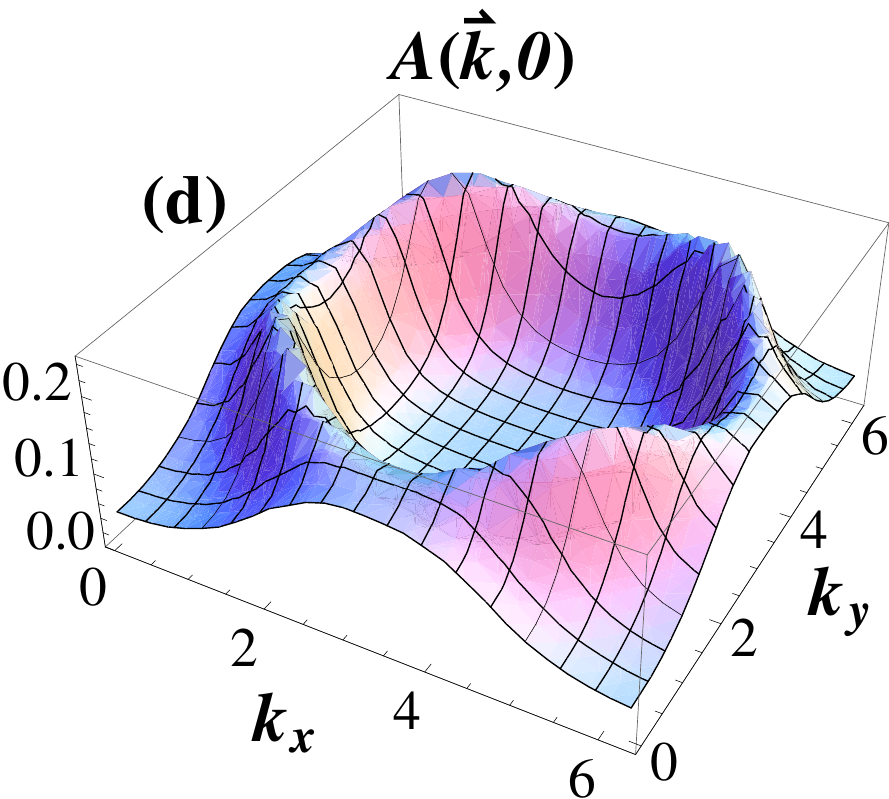}
\end{tabular}
\caption{Temperature dependence of the spectral function at the Fermi energy $A(\mathbf{k},\omega=0)$ with $\mathbf{k}$
in the first Brillouin zone ($V=2.4$): From (a) to (d) $T=0.03, 0.09, 0.15, 0.3$. At $T=0.03$, four sharp spectal peaks at four nodes; With the increasing temperature, $A(\mathbf{k},0)$ decreases at the nodes while piles up at the other $\mathbf{k}$'s of the underlying Fermi surface;
At $T=0.3$, the spectral distribution seems like a bowl. The gap of the antinode survives at the high temperature (For $28\times 28$ lattice, $T>0.3$; For smaller size lattice, we find it dose not close even at $T>0.5$).}
\label{spec3D}
\end{figure}

The single-particle excitation spectra of electrons moving in a fluctuating pairing field is highly nontrivial.
Figure~\ref{spec3D} shows the evolution of $A(k,\omega=0)$ as a function of temperature. The momentum $\mathbf{k}$ is
in the first Brillouin zone and the energy $\omega=0$ on the Fermi level. We set the chemical potential $\mu=-0.83$
intentionally to have more discrete momenta on the underlying Fermi surface and the resulting electron number is around $0.9$.
At $T=0.03$, four sharp spectral peaks right at the four gap nodes are clearly resolved in Fig.~\ref{spec3D}(a), which indicates that at temperatures well below $T_\text{KT}$ the pair fluctuations are rather weak and the Fermi surface are actually point like as in pure
$d$-wave superconductors.
At $T=0.09$, the height of the peaks falls while their profile extends towards the antinodal direction, i.e.~the
spectral weight of other $\mathbf{k}$ points along the underlying Fermi surface increases. From Figure~\ref{spec3D},
we find that this pile-up effect of spectral weight at the vicinity of the underlying Fermi surface increases with
temperature, which is consistent with the ARPES observations \cite{kanigel06,kanigel07} as well as the theoretical picture \cite{berg}.

\begin{figure*}[htb]
\includegraphics[width=5cm]{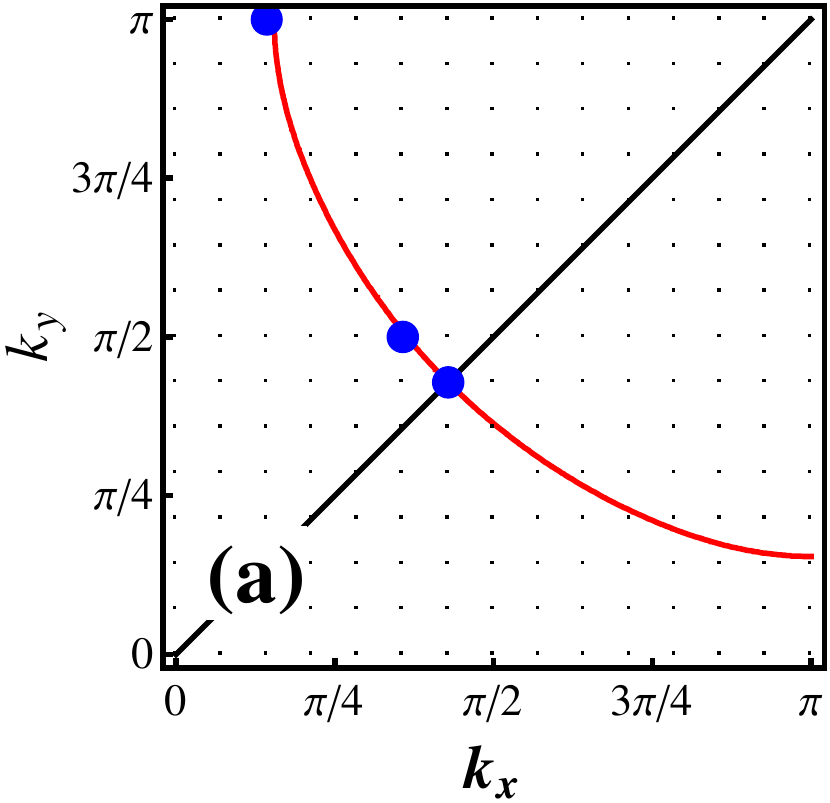}
\includegraphics[width=5cm]{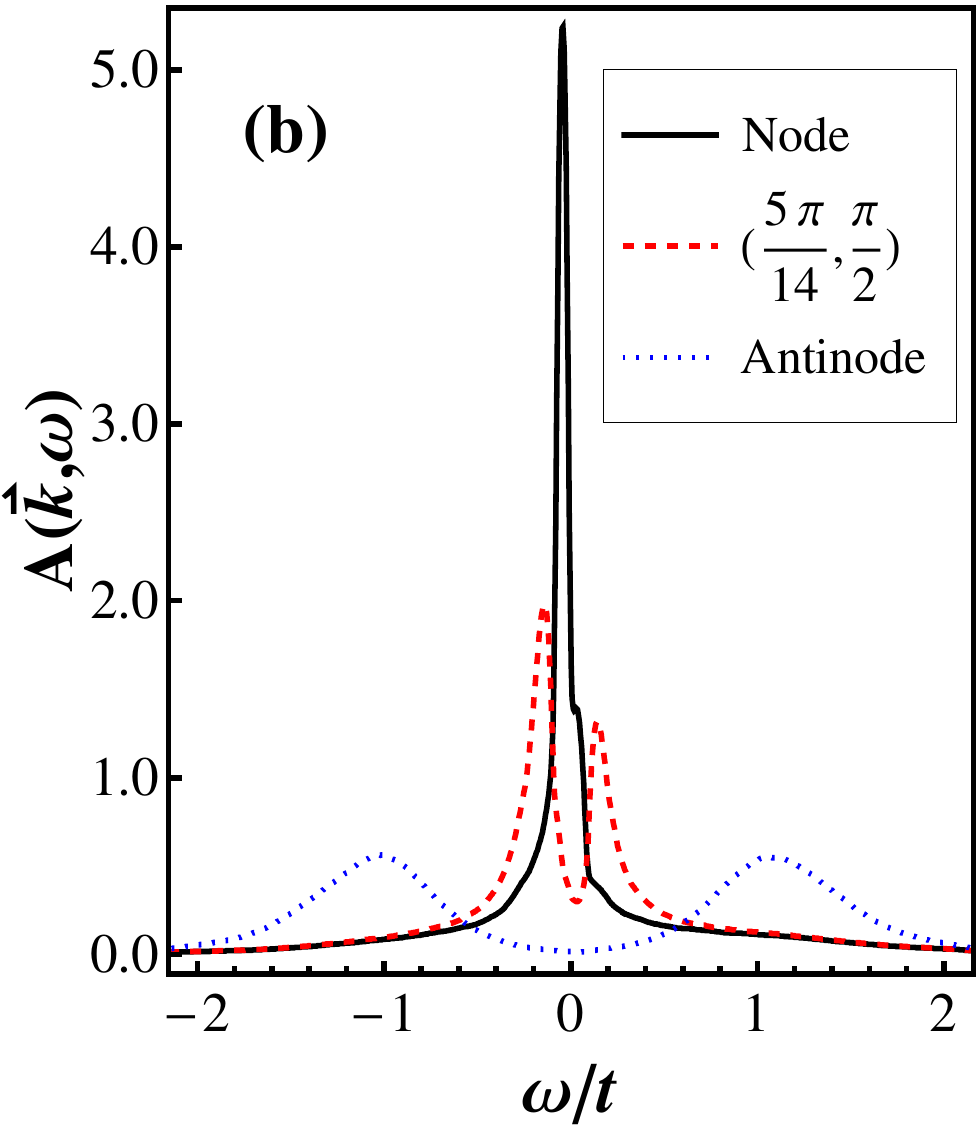}
\includegraphics[width=5cm]{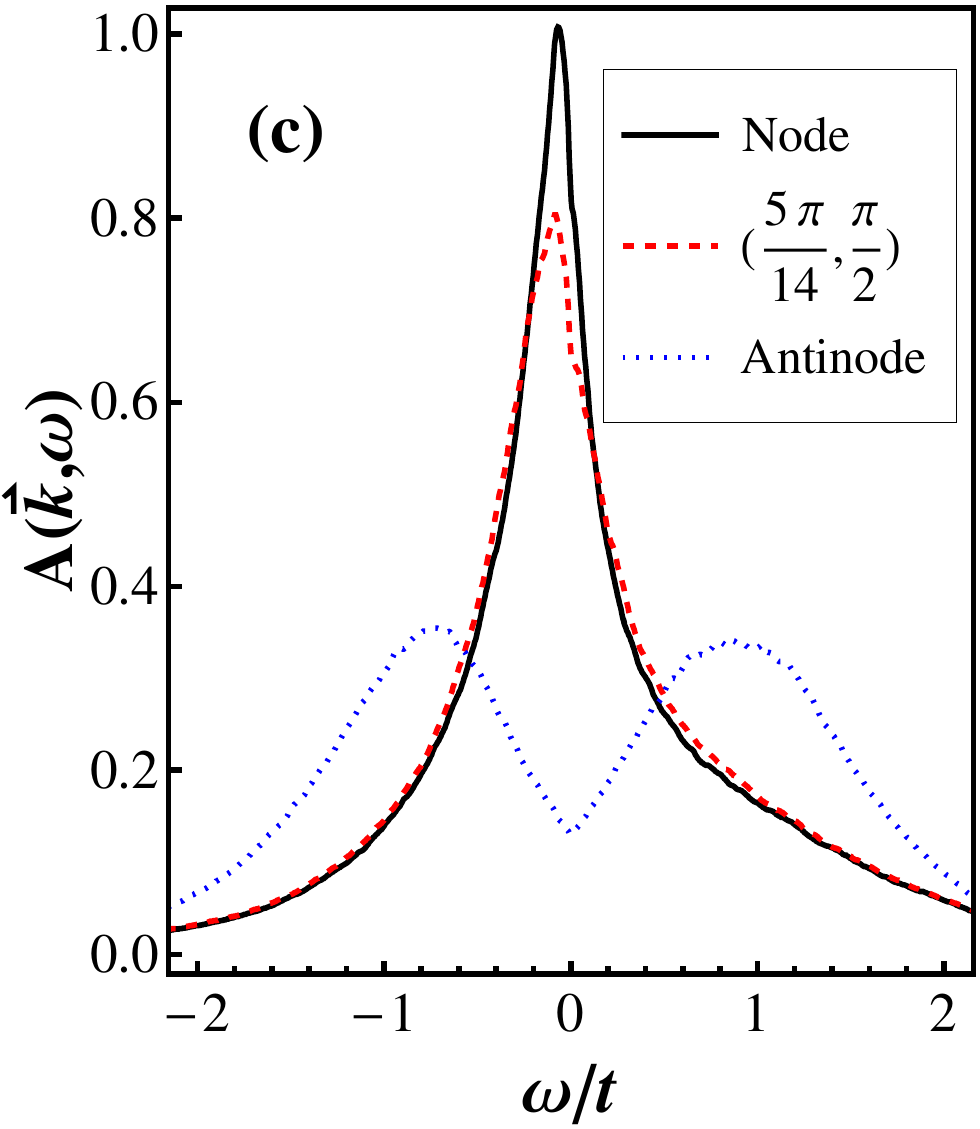} \\
\includegraphics[width=5cm]{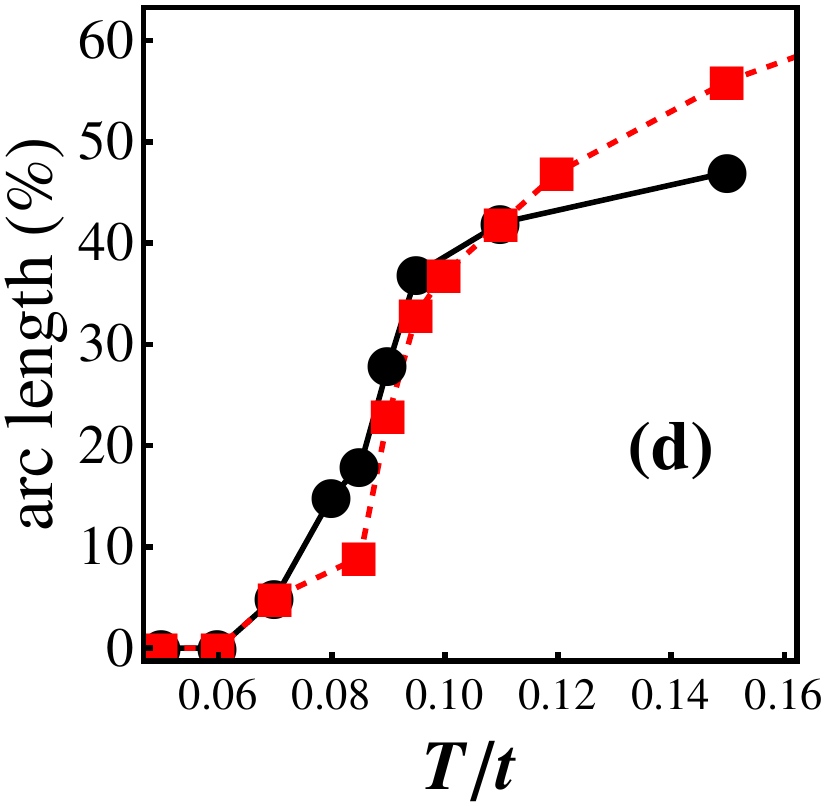}
\includegraphics[width=5cm]{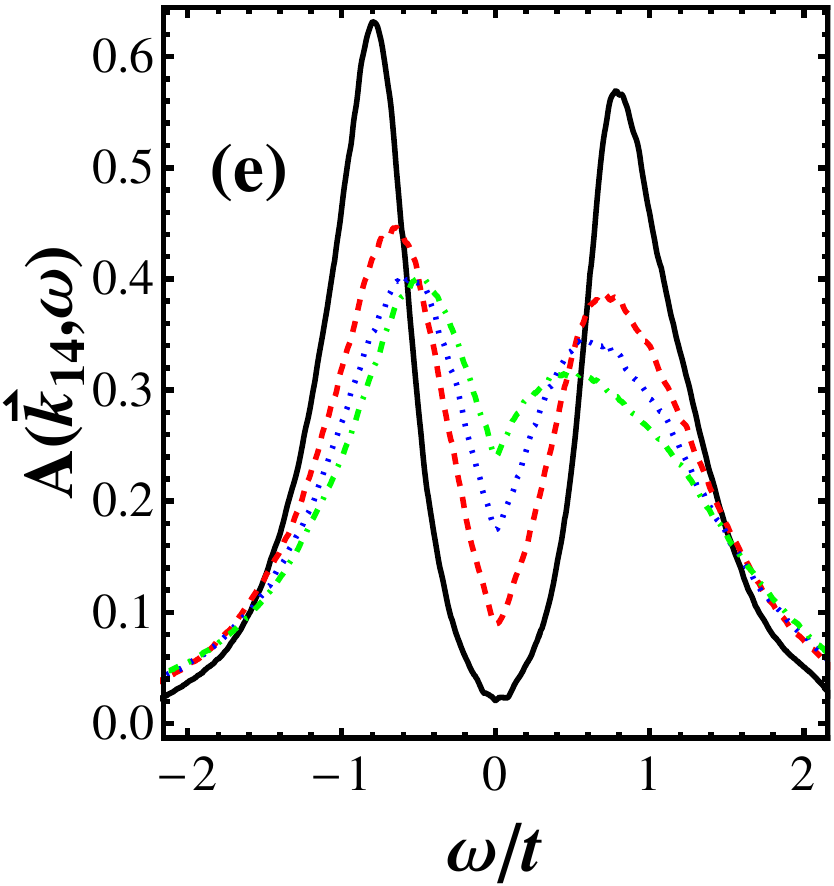}
\includegraphics[width=5cm]{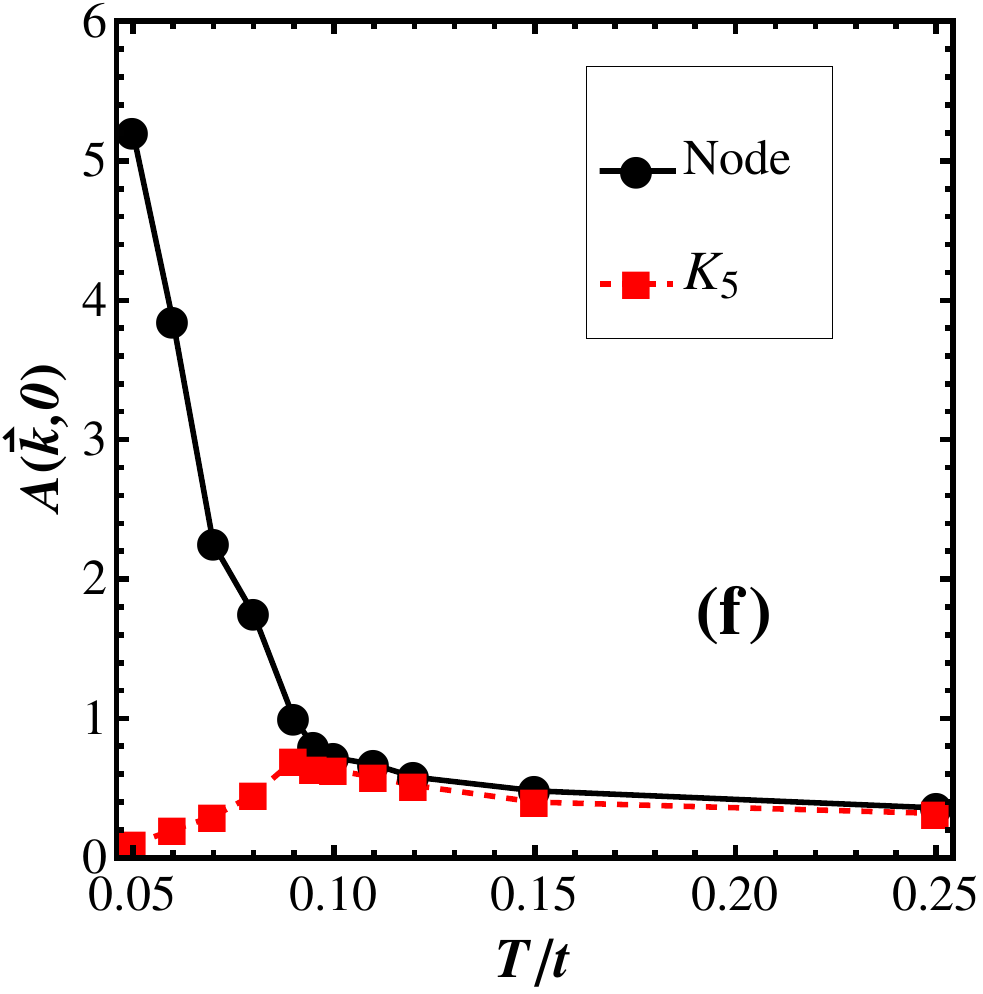}
\caption{ (a), The Fermi surface shown as red line in the first quarter of the Brillouin zone. Three selected $\mathbf{k}$ points lying on the Fermi surface are shown as large blue dots. The spectral function $A(\mathbf{k},\omega)$ as a function of $\omega$ for this three $\mathbf{k}$ points with (b) $T=0.05$ and (c) $T=0.09$. (d) The Fermi-arc length versus temperature. (e) Temperature dependence of $A(\mathbf{k}=\mathbf{k}_{14},\omega)$. Black solid line($T=0.05$), red dashed line($T=0.08$), blue dotted line($T=0.09$), and green dot-dashed line($T=0.1$). (f), $A(\mathbf{k},\omega=0)$ as a function of temperature for the node point and $\mathbf{k}_{5}$. For all panels $V=2.4$. }
\label{arclength}
\end{figure*}
Next, we show the energy distribution of the spectral function in Fig.~\ref{arclength}(b) and (c) to examine the spectral gaps opened at different $\mathbf{k}$'s on the underlying Fermi surface at two different temperatures. The selected three $\mathbf{k}$'s are the node $(\frac{6\pi}{14},\frac{6\pi}{14})$, the wave vector $(\frac{5\pi}{14},\frac{\pi}{2})$ near the node and the antinode $(\frac{2}{14\pi},\pi)$ as shown in Fig.~\ref{arclength}(a). At $T=0.05$ below
the KT transition, $A(\mathbf{k},\omega)$ for $\mathbf{k}$ at node displays a sharp peak located at zero energy. Away from the node,
we find that the spectral gap opens at the selected momentum $\mathbf{k}=(\frac{5\pi}{14}, \frac{\pi}{2})$ closest to the node and increases to its largest value at the antinode, which conforms to the characteristic of $d$-wave superconducting gap function. At higher temperature $T=0.09$,
we find that the spectral peak at the node is lowered in consistent with the observation of Fig.~\ref{spec3D}(b).
Moreover the spectral gap at $\mathbf{k}=(\frac{5\pi}{14}, \frac{\pi}{2})$ is closed, while a spectral peak is piled up at
the zero energy, which signals the formation of the so-called Fermi arc that extends from node as far as to
$\mathbf{k}=(\frac{5\pi}{14}, \frac{\pi}{2})$.
To quantify the length of the Fermi arc, we examine the loss of the spectral weight~\cite{kanigel06}
due to the opening of the spectral gap $L=[1-\frac{2A(\mathbf{k},\omega=0)}{A(\mathbf{k},\omega=-\Delta)+A(\mathbf{k},
\omega=\Delta)}]$, where the spectral gap $\Delta$ is measured as half the peak to peak separation of the spectral function. In the ideal case $L=1$ means the opening of a full spectral gap; while $L=0$ identifies the closing of the gap, or in other words the formation of the Fermi arc. Here in analyzing our numerical results, we set both $L=0.1$ and $L=0.15$ as the threshold for the arc formation.
The results are plotted in Fig.~\ref{arclength}(d) where the variation of the length of Fermi arcs as a function
of temperature is shown. For both parameters, the arc length exhibits apparent rise near $T_{c}$, which is consistent
with the APRES measurement \cite{kanigel07}. In addition, this jump locates around the same temperature where SFD,
phase correlation, and the vortex density changes most remarkably, indicating the importance of pair fluctuation
in the formation of the Fermi arc. We cut the underlying Fermi surface between the node and antinode in the first
quarter of the first Brillouin zone into $20$ equally spaced parts, with $\mathbf{k}_{0}$ denoting the node and $\mathbf{k}_{20}$
the antinode. We examine the $\mathbf{k}_{14}$ point in Fig.~\ref{arclength}(e). It is clearly shown that there is a continuous
increase of the spectral weight at the Fermi level from $T=0.05$ to $T=0.1$, while the spectral gap shrinks as
the temperature increases, which can be explained by the increasing broadening effect due to the thermal pairing fluctuation.

Now we report the second shift of the zero energy spectral weight. According to Figs.~\ref{arclength}(b) and (c), we notice that the zero energy spectral weight at the node/antinode is always decreaing/increasing with temperature.
However for the $\mathbf{k}$ points near the node, the zero energy spectral weight first increases with temperature and then decreases above a temperature whose value depends on $\mathbf{k}$ as shown in Fig.~\ref{arclength}(f). We call this phenomenon the second shift, which can be understood according to the theory of Berg and Altman~\cite{berg}.
Because of the Doppler-shift effect, the zero-energy spectral weight of the node is transferred to its neighboring $\mathbf{k}$ points and is gradually exhausted nearby $T_{c}$; the neighboring points received the zero energy spectral weight from the node, and also due to the same effect, shift their zero-energy spectral weight to their neighboring points.
The higher the temperature is, the less they get and the more they shift. For high enough temperature, both the node and the neighboring points have approximately equal amount of the zero-energy spectral weight as shown in figure~\ref{arclength}(f), and the zero-energy spectral weights of these points saturate and begin to decrease.

\section{Conclusions}
\label{conclusion}

In conclusion, we have carried out the classical Monte Carlo simulation of the 2D attractive Hubbard model.
We have presented a local-update procedure based on the Matsubara Green's Function using Nambu-Gor'kov formalism,
which speeds up the exploring of the configuration space. We found that thermal fluctuations do contribution to
the pile up of low-energy spectral weight on the underlying Fermi surface and the evolution of fermi arcs with temperature.
The abrupt jump of the arc length is a qualitative result caused by the continuous piling up of zero energy spectral weight,
and the second shift suggests that E. Berg and E. Altman's idea works better than simply thermally broadening effect.
Finally, taking superfluid density as the SC criteria is effective when interaction $V$ is small.

The work was supported by the Natural Science Foundation of China No.10674179.

\section{Appendix: superfluid density}

The superfluid density can be obtained applying the linear response theory to the homogenous superconducting
state~\cite{scalapino}. Here we will use Gor'kov Green's function to express the superfluid weight.
We start from the following formula ~\cite{scalapino}
\begin{equation}
\frac{D_s}{\pi e^2} = \frac{1}{N\hbar^2}\langle - \hat{K}_x \rangle - \Pi_{xx}(q_x=0,q_y\rightarrow 0, i\Omega_m=0), \label{ds}
\end{equation}
where $D_s$ represents the superfluid weight and measures the ratio of superfluid density to mass.
Here $\hat{K}_x$ denotes the electron kinetic energy along $x$ direction and
\begin{equation}
\langle - \hat{K}_x\rangle = \sum_{\mathbf{k}\sigma} \frac{\langle c_{\mathbf{k}\sigma}^\dagger c_{\mathbf{k}\sigma} \rangle}{m_\mathbf{k}^x}
= 2 \sum_{\mathbf{k},\omega_n} \frac{G_{11}(\mathbf{k},i\omega_n) e^{i\omega_n 0^+} }{m_\mathbf{k}^x}, \label{kinetic}
\end{equation}
$m_\mathbf{k}^x = (\partial^2 \varepsilon_\mathbf{k}/\partial k_x^2)^{-1}$ the electron effective mass with
$\varepsilon_\mathbf{k}=-2t(\cos k_x + \cos k_y) -4t^\prime \cos k_x\cos k_y$ the electron dispersion.
$\Pi_{xx}$ is the current-current correlation function defined in momentum and imaginary-time space
\begin{equation}
\Pi_{xx}(\mathbf{q},\tau) = \frac{1}{N} \langle \text{T}_\tau \hat{j}_x(\mathbf{q},\tau) \hat{j}_x(-\mathbf{q},0) \rangle,
\end{equation}
where $\hat{j}_x(\mathbf{q}) = \frac{1}{\hbar}e^{iq_x/2} \sum_{\mathbf{k}\sigma}
v_\mathbf{k+q/2}^x c_{\mathbf{k}\sigma}^\dagger c_{\mathbf{k+q}\sigma}$ the current-density operator, and $v_\mathbf{k}^x = \partial\varepsilon_\mathbf{k}/\partial k_x$ denotes the group velocity of electron. Performing Fourier transform with respect to imaginary time, we have
\begin{equation}
\Pi_{xx}(\mathbf{q},i\Omega_m) =\int_0^\beta d\tau e^{i\Omega_m \tau} \Pi_{xx}(\mathbf{q},\tau), \label{corrfunc}
\end{equation}
Combining Eq.~(\ref{ds}), (\ref{kinetic}) and (\ref{corrfunc}) followed by straightforward derivation,
we have the equation for superfluid density expressed using the Gor'kov Green's function,
\begin{widetext}
\begin{eqnarray}
\frac{n_s}{m^*}\equiv\frac{D_s}{\pi e^2} = \frac{2}{N\hbar^2} \sum_{\mathbf{k},\omega_n} \frac{G_{11}(\mathbf{k},i\omega_n) e^{i\omega_n 0^+} }{m_\mathbf{k}^x} +
\left.
\frac{1}{N \hbar^2\beta}\sum_{\mathbf{k},\omega_n}
\left(v_\mathbf{k+q/2}^x\right)^2
\text{tr} [ G(\mathbf{k},i\omega_n) G(\mathbf{k}+\mathbf{q},i\omega_n) ]
\right|_{q_x=0,q_y\rightarrow0} . \label{dsgreen}
\end{eqnarray}
\end{widetext}

Then we will use the above formula to pair-fluctuating superconductors, whose superfluid weight is given by
\begin{equation}
D_s = \frac{ \int D\Delta D\bar{\Delta} e^{-\beta\Omega(\Delta,\bar{\Delta})} D_s(\Delta,\bar{\Delta}) }
{\int D\Delta D\bar{\Delta} e^{-\beta\Omega(\Delta,\bar{\Delta})}}, \label{dsavg}
\end{equation}
where, $D_s(\Delta,\bar{\Delta})$ denotes the superfluid weight for a certain configuration $\Delta$. Considering that $\Delta$ is
spatially inhomogeneous, we should first perform Fourier transform on the real-space Gor'kov Green's
function which has been obtained and updated
during the random walk through the configuration space,
\begin{equation}
G(\mathbf{k},i\omega_n;\Delta) = \frac{1}{N} \sum_{i,j} G(i,j,i\omega_n;\Delta) e^{i\mathbf{k}\cdot(\mathbf{i-j})}. \label{ft}
\end{equation}
After the transformation, Eq.~(\ref{ft}) is inserted into Eq.~(\ref{dsgreen}) to calculate the superfluid density corresponding to
one configuration and then using Eq.~(\ref{dsavg}) to obtain the statistical average over configurations.

\end{document}